\documentclass[a4paper,12pt]{article}
\usepackage{amsmath}
\usepackage{amssymb}
\usepackage{braket}
\usepackage{cancel}
\usepackage{graphicx}
\usepackage{color}
\usepackage[utf8]{inputenc}
\usepackage{hyperref}
\usepackage{multirow}
\usepackage{slashed}
\usepackage{booktabs}
\usepackage[normalem]{ulem}
\usepackage{wasysym}
\usepackage{amssymb}
\usepackage{braket}
\usepackage{simplewick}
\usepackage{array}
\usepackage{graphicx}
\usepackage{cite}
\usepackage{float}
\usepackage{titlesec}
\usepackage{listings}
\usepackage{hyperref}
\usepackage{bbold}
\usepackage[english]{babel}
\usepackage{xcolor}
\usepackage{subcaption}
\usepackage{braket}
\usepackage{comment}
\usepackage{enumitem}
\usepackage{bm}
\usepackage[export]{adjustbox}
\allowdisplaybreaks

\usepackage{lscape}

\usepackage[english]{babel}

\usepackage{graphicx}

\usepackage{geometry}
\renewcommand{\d}{\mathrm{d}}
\newcommand{\pert}{{\mathrm{pert}}}
\newcommand{\kin}{{\mathrm{kin}}}
\newcommand{\OS}{{\mathrm{OS}}}
\newcommand{\GeV}{{\mathrm{GeV}}}
\newcommand{\api}{\left( \frac{\alpha_s}{\pi} \right)}
\begin{document}

\begin{titlepage}

\begin{flushright}
{\small
CERN-TH-2024-025\\
ZU-TH 14/24
}
\end{flushright}

\vskip1cm
\begin{center}
{\Large \bf\boldmath 
NNLO QCD corrections to the $q^2$ spectrum of inclusive
semileptonic $B$-meson decays
}
\end{center}

\vspace{0.5cm}
\begin{center}
{\sc Matteo Fael$^a$ and Florian Herren$^b$} \\[6mm]

{\it $^a$ Theoretical Physics Department, CERN, 1211 Geneva, Switzerland}\\[0.3cm]

{\it $^b$ Physics Institute, Universität Zürich, Winterthurerstrasse 190, CH-8057 Zürich, Switzerland}\\[0.3cm]
\end{center}

\vspace{0.6cm}
\begin{abstract}
\vskip0.2cm\noindent
We calculate the next-to-next-to-leading order QCD corrections to the leptonic invariant
mass ($q^2$) spectrum  of semileptonic $b \to c$ inclusive decays, taking into account 
the mass of the charm quark and the charged lepton in the final state.
We obtain analytic results in terms of generalized polylogarithms and 
present numerical studies of the $\mathcal{O}(\alpha_s^2)$ corrections to the $q^2$ spectrum of 
$b \to c \ell \bar \nu_\ell$ decays, for $\ell =e, \mu$ and $\tau$, in the kinetic scheme.
Our computation can be used to incorporate the recent measurements of $q^2$ moments by Belle and Belle II into global fits of inclusive semileptonic $B$-decays.
\end{abstract}

\end{titlepage}

\section{Introduction}
The Heavy Quark Expansion (HQE) has become a pillar in the theoretical description
of inclusive decays of heavy hadrons, allowing the derivation of precise predictions 
with reliable estimates of the uncertainties. 
One of its main applications is the study of the inclusive decay $B \to X_c \ell \bar \nu_\ell$. 
Thanks to its relatively large rate and clean experimental signature, studies of inclusive semileptonic decays have lead to precise determinations of the magnitude of the Cabibbo-Kobayashi-Maskawa matrix element $V_{cb}$. 

The HQE allows to describe both the total decay rate and various kinematic distributions
as a double series expansions in the strong coupling constant $\alpha_s$ and $\Lambda_\mathrm{QCD}/m_b$.
Various measurements of moments of the charged lepton energy and the hadronic invariant mass in
$B \to X_c \ell \bar \nu_\ell$ decays have been performed by BABAR \cite{BaBar:2004bij,BaBar:2009zpz}, BELLE \cite{Belle:2006kgy,Belle:2006jtu}, CLEO \cite{CLEO:2004bqt}, CDF \cite{CDF:2005xlh} and DELPHI \cite{DELPHI:2005mot}.
The comparison of experimental measurements with the predictions calculated within the HQE
has lead to determinations of $|V_{cb}|$ with a $1.1\%$ accuracy~\cite{Alberti:2014yda,Bordone:2021oof,Finauri:2023kte} 
and the so called ``HQE parameters'', the non-perturbative matrix elements, such as $\mu_\pi^2, \mu_G^2, \rho_D^3$ and $\rho_{LS}$.

The HQE parameters are important theoretical inputs not only for the extraction
of $|V_{cb}|$, but also for the extraction of $|V_{ub}| $ from $B \to X_u \ell \bar \nu_\ell$ decays
since the moments of the shape functions can be 
related to the HQE parameters extracted in $b \to c$ decays. 
Moreover predictions for other kinds of processes require precise knowledge of the HQE parameters, as for instance the $B$-meson lifetimes~\cite{Lenz:2022rbq,Albrecht:2024oyn} and the 
rare decay $B \to X_{d,s} \ell \bar \ell$~\cite{Huber:2019iqf,Huber:2020vup}.

An alternative method for the determination of $|V_{cb}|$ has been proposed in Ref.~\cite{Fael:2018vsp}
and is based on the measurement of the leptonic invariant mass ($q^2$) spectrum and the
branching ratio as a function of a lower cut on $q^2$. These observables are invariant under reparametrization, 
a symmetry within the HQE reflecting Lorentz invariance of the underlying QCD.
Reparametrization invariant (RPI) quantities depend on a reduced set of HQE parameters~\cite{Mannel:2018mqv,Fael:2018vsp}.
The smaller set of parameters necessary in a global fit of these observables (eight instead of 13 up to $1/m_b^4$)
enabled to extract $|V_{cb}|$ and the HQE parameters up to $1/m_b^4$
in a completely data-driven way~\cite{Bernlochner:2022ucr}, based on recent measurements 
of $q^2$ moments by Belle~\cite{Belle:2021idw} and Belle-II~\cite{Belle-II:2022evt}.
The new measurements of the $q^2$ moments have also been included in a global 
fit of $|V_{cb}|$~\cite{Finauri:2023kte}, together with the lepton energy and $M_X$ moments, 
finding that the new data are compatible with the other measurements and slightly 
decreasing the uncertainty on the HQE parameters and on $|V_{cb}|$.
Recently, the RPI operator basis has been extended up to order $1/m_b^5$~\cite{Mannel:2023yqf}.

Given the precision achieved by experimental measurements, which show a percent or even sub-percent
relative accuracy for certain observables, a good control of 
perturbative and non-perturbative effects in the HQE is mandatory, also in light of the
rather large value of the strong coupling constant $\alpha_s(m_b) \simeq 0.22$.
At the partonic level, next-to-leading order (NLO) corrections are available from Ref.~\cite{Jezabek:1988iv,Jezabek:1996db,Aquila:2005hq}. 
The $\mathcal{O}(\alpha_s)$ triple differential distribution up to the power-suppressed terms $\mu_\pi^2$
and $\mu_G^2$ were presented in Ref.~\cite{Becher:2007tk,Alberti:2012dn,Alberti:2013kxa},
while NLO corrections proportional to $\rho_D$ are available only for the total rate and the $q^2$ spectrum~\cite{Mannel:2021zzr}.

The $\mathcal{O}(\alpha_s^2)$ corrections to the free quark decay $b \to X_c \ell \bar \nu_\ell$ 
are also required for a consistent theoretical description and to match
the experimental accuracy. The next-to-next-to-leading order (NNLO) corrections 
to the hadronic invariant mass and charged-lepton energy moments have been calculated 
in~\cite{Melnikov:2008qs,Biswas:2009rb}.
No result is available for the $q^2$ spectrum for arbitrary values of $q^2$,
only for $q^2=0$~\cite{Czarnecki:1997hc}, $q^2=(m_b-m_c)^2$~\cite{Czarnecki:1997cf,Czarnecki:1996gu} 
and $q^2=m_c^2$~\cite{Czarnecki:1998kt}. 
While the calculations in these three special points allowed the authors of Ref.~\cite{Czarnecki:1998kt}
to estimate the non-BLM corrections at $\mathcal{O}(\alpha_s^2)$ with a relative $30\%$ uncertainty,
their result is unsuited to calculate higher moments of the $q^2$ spectrum
with sufficient precision.\footnote{We contacted the authors of Ref.~\cite{Biswas:2009rb},
however they could not retrieve their original Monte Carlo code.}
Analytic expressions up to $\mathcal{O}(\alpha_s^3)$ for the $q^2$ moments without threshold selection 
cuts have been presented in Ref.~\cite{Fael:2022frj}, while Ref.~\cite{Finauri:2023kte} 
presented an evaluation of the $\alpha_s^2 \beta_0$ corrections, 
utilizing the BLM correction to the triple differential rate from Ref.~\cite{Aquila:2005hq}.

The goal of this paper is to present the complete NNLO QCD corrections to the $q^2$ spectrum
of $b \to X_c \ell \bar \nu_\ell$. 
At variance with the numerical approach used in~\cite{Melnikov:2008qs,Biswas:2009rb},
based on sector decomposition, recent developments in analytic approaches to multi-loop 
computations allow us to calculate the differential rate w.r.t.\ $q^2$ in an analytic form
and write it in terms of generalized polylogarithms (GPLs)~\cite{Goncharov:1998kja,Goncharov:2001iea}.
Our results can be used to calculate the NNLO corrections to the $q^2$ moments
with arbitrary cuts on $q^2$. The inclusion of the results presented in this paper
into global fits will allow to better assess the theoretical uncertainty in the 
prediction.

The paper is organized as follows. In Section~\ref{sec::details} we present the 
details of the calculation, in particular we discuss how we obtain analytic results 
for the three-loop master integrals at NNLO. Section~\ref{sec:results} presents
our numerical results for the differential rate and the moments in the on-shell scheme
and the kinetic scheme. We will discuss also the impact on the decay $B \to X_c \tau \bar \nu_\tau$. We conclude in Section~\ref{sec:conclusions}.

\section{Details of the calculation}
\label{sec::details}
Let us now discuss the details of the calculation.
We consider the semileptonic decay of a bottom quark mediated by the weak interaction
\begin{equation}
    b (p_b) \to X_c(p_X) \ell(p_\ell) \bar \nu_\ell(p_\nu),
    \quad \text{with} \quad \ell = e,\mu,\tau,
\end{equation}
where $X_c$ generically denotes a state containing a charm quark, plus additional gluons and/or quarks.
The mass of the charged lepton $\ell$ is denoted by $m_\ell$ while the neutrino is considered massless.
The masses of the bottom and charm quark are $m_b$ and $m_c$, respectively, and we
introduce their ratio $\rho =m_c/m_b$.
In the following we study the spectrum of the leptonic invariant mass $q^2 = p_L^2$ with $p_L = p_\ell + p_\nu$.
We begin by writing the differential rate w.r.t.\ $q^2$ as 
\begin{equation}
    \frac{d\Gamma}{d \hat q^2} =
    \frac{G_F^2 m_b^5}{192 \pi^3} |V_{cb}|^2
    \Bigg[
    F_0(\rho,\hat q^2)
    + \frac{\alpha_s}{\pi}
    F_1(\rho,\hat q^2)
    + \left(
     \frac{\alpha_s}{\pi}
    \right)^2
    F_2(\rho,\hat q^2)
    \Bigg]
    +O\left( \frac{1}{m_b^2}\right),
    \label{eqn:decayrate}
\end{equation}
where $\hat q^2 = q^2/m_b^2$ and $F_i$ stands for the differential decay rate at leading, next-to-leading and next-to-next-to-leading order, respectively. 
The functions $F_0$ and $F_1$ are known since a long time~\cite{Jezabek:1988iv}.
For the power corrections up to $1/m_b^3$ the NLO corrections have been presented in Refs.~\cite{Mannel:2021ubk,Mannel:2021zzr,Moreno:2022goo}.
The quark masses $m_b$ and $m_c$ are renormalized in the on-shell scheme. 
The strong coupling constant $\alpha_s = \alpha_s^{(5)}(\mu_s)$ is renormalized in the $\overline{\mathrm{MS}}$ scheme
with five active flavours with $\mu_s$ being the renormalization scale.

\begin{figure}
    \centering
    \includegraphics[width=0.8\textwidth]{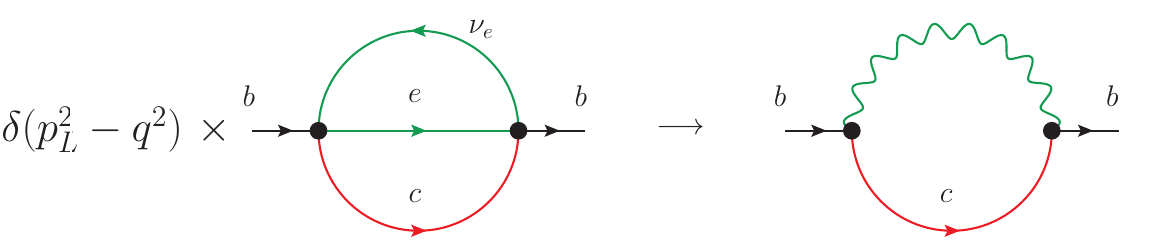}
    \caption{Phase space intergration in the presence of the constraint on the leptoinc invariant mass $\delta(p_L^2-q^2)$ 
    is mapped into the cuts of forward $b \to b$ scattering amplitude where the two leptons are
    replaced by a fake spin-1 massive particle.}
    \label{fig:fake_particle}
\end{figure}

In order to calculate the NNLO QCD corrections to the $q^2$ spectrum, we follow the method of Refs.~\cite{Mannel:2021ubk,Mannel:2021zzr}.
The idea is to consider the differential rate in the presence of a constraint on $q^2$.
The phase-space decomposition suitable for $b \to X_c \ell \bar \nu_\ell$ is carried out by assuming
a sequence of two two-body decays. First, the bottom quark decays into an off-shell $W$-boson
and the charm quark, then the virtual $W$-boson decays into the lepton and the neutrino:
\begin{equation}
    d\Gamma = 
    \frac{(2\pi)^d}{2m_b}
    \delta(p_L^2-q^2)
    W_{\mu\nu} L^{\mu\nu} 
    d\Phi_2 (p_b; p_L, p_X)
    d\Phi_2(p_L; p_\ell, p_\nu)
    (2\pi)^{d-1} dq^2 
\end{equation}
where the integration is performed in a $d$-dimensional space, with $d=4-2\epsilon$.
$W^{\mu\nu}$ and $L^{\mu\nu}$ are the hadronic and leptonic tensors.
The element of $n$-body phase space is given by
\begin{equation}
    d\Phi_n(P; p_1, \dots, p_n) = 
    \delta^{(d)}\left(P-\sum_{i=1}^n p_i\right) \prod_{i=1}^n 
    \frac{d^{d-1} p_i}{(2\pi)^{d-1} 2E_i}.
\end{equation}
The constraint $\delta(p_L^2-q^2)$ enforces the dilepton system to have an invariant mass equal to $q^2$.
Since the hadronic tensor depends only on $p_L$ and $p_b$, we can integrate the leptonic tensor with respect to 
the phase-space of charged lepton and neutrino~\cite{Moreno:2022goo}:
\begin{align}
    \mathcal{L}^{\mu\nu} (p_L) &= 
    \int L^{\mu\nu} d\Phi_2(p_L; p_e, p_\nu) 
    \notag \\ &
    = 
    \frac{1}{384 \pi^5} 
    \left( 
    1-\frac{m_\ell^2}{p_L^2}
    \right)^2
    \left[
    \left( 
    1+\frac{2 m_\ell^2}{p_L^2}
    \right)
    p_L^\mu p_L^\nu 
    - g^{\mu\nu}p_L^2 
    \left( 
    1+\frac{m_\ell^2}{2p_L^2}
    \right)
    \right]
    +\mathcal{O}(\epsilon).
    \label{eq::Lcal}
\end{align}
Higher order terms in $\epsilon$ are not necessary since the leptonic tensor does not 
enter into the renormalization, i.e.\ $\mathcal{L}^{\mu\nu}$ is always contracted 
with the renormalized hadronic tensor. 
After inserting Eq.~\eqref{eq::Lcal} into the differential rate formula, 
we represent the $\delta$ function as the imaginary part of a propagator:
\begin{equation}
    \delta(p_L^2 - q^2) \to
    \frac{1}{2\pi i}\left[ 
    \frac{1}{p_L^2-q^2-i 0}
    -\frac{1}{p_L^2-q^2+i 0}
    \right],
\end{equation}
In other words, we treat $\delta(p_L^2 - q^2)$ as an on-shell condition for a ``fake'' particle. 
Next we apply the reverse unitarity method~\cite{Anastasiou:2002yz} and map the calculation of the various interference terms 
integrated over the final state phase-space into the evaluation of ``cuts'' of forward $b \to b$ scattering amplitudes.
In this way, the differential rate can be obtained from the the imaginary part of a $b \to b$ two-point amplitude,
where the two leptons with constrained invariant mass are replaced by a fake (colorless) spin-1 particle 
with mass equal to $q^2$ (see Fig.~\ref{fig:fake_particle}).
Note that with this method, we need to consider loop diagrams with one loop less compared to the original diagrams, i.e.\ to calculate the $q^2$ spectrum at LO, NLO and NNLO we have to consider one-, two- and three-loop diagrams. One the other hand, the Feynman integrals now depend on two dimensionless variables: $\rho$ and $\hat q^2$.

Let us now discuss the technical details of the calculation.
We generate with \texttt{qgraf}~\cite{Nogueira:1991ex} one, two and three loop diagrams (as the ones on the r.h.s.\ of Fig.~\ref{fig:fake_particle}) and use \texttt{Tapir}~\cite{Gerlach:2022qnc} to map each diagram to a predefined integral family. We use the program \texttt{exp}~\cite{Seidensticker:1999bb} to rewrite
the output to \texttt{FORM}~\cite{Kuipers:2012rf} notation.
In this way we express the three-loop $b \to b$ amplitude as linear combinations 
of scalar Feynman integrals with nine indices,
where eight correspond to the exponents of propagators
and the remaining one to the exponent of an irreducible numerator. 
In total we have 21 integral families at three loops.

Before performing the IBP reduction, we find with a basis of master integrals such that the denominators in the reduction tables completely 
factorize into polynomials depending either on $\rho$ and $\hat q^2$ or $d$ (see Ref.~\cite{Smirnov:2020quc,Usovitsch:2020jrk}).
To construct this basis, we first reduce a set of seed integrals up to two dots and one scalar product for every integral family individually with the help of \texttt{Kira}~\cite{Maierhofer:2017gsa,Klappert:2020nbg} and \texttt{Fermat}~\cite{fermat}. 
As initial basis we simply take the default master integrals.
These reduction tables then serve as input to search for a good basis
for every family with the help \texttt{ImproveMasters.m} developed in
Ref.~\cite{Smirnov:2020quc}.

The IBP reduction of the integrals appearing in the amplitude is then performed with \texttt{Kira}. First, we reduce the integrals for every family individually to the good basis of this family. Then we employ symmetries between the families to reduce the number of master integrals. 
Afterwards we identify the master integrals which have an imaginary part
while setting to zero those which are purely real, e.g.\ tadpole integrals.
We find one, six and 98 master integrals at one, two and three loops.

We solve the master integrals in an analytic way by using the method of differential equations~\cite{Kotikov:1990kg,Gehrmann:1999as}.
We establish a set of differential equations by differentiating the 98 master integrals with respect to $\rho$ and $\hat q^2$ and reducing the resulting integrals again to master integrals with \texttt{Kira}. We obtain a system of the form
\begin{align}
    \frac{\partial \bm J}{\partial \rho} &=
    \mathbb{M}_\rho(\rho,\hat q^2,\epsilon) \bm J, &
    \frac{\partial \bm J}{\partial \hat q^2} &=
    \mathbb{M}_{q^2}(\rho,\hat q^2,\epsilon) \bm J ,
\end{align}
where $\bm J = (J_1, \dots, J_{98})$ is the set of master integrals, $\epsilon$ is the dimensional regularization parameter
while $\mathbb{M}_\rho$ and $\mathbb{M}_{q^2}$ are $98 \times 98$ matrices, rational in $\rho$, $\hat q^2$ and $\epsilon$.

\begin{figure}
    \centering
     \begin{subfigure}[c]{0.3\textwidth}
         \centering
         \includegraphics[width=\textwidth,trim=2cm 0 2cm 0, clip]{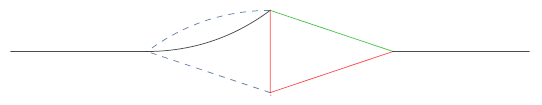}
         \caption{Class I}
         \label{fig:sampleI}
    \end{subfigure}
    \quad
    \begin{subfigure}[c]{0.3\textwidth}
         \centering
         \includegraphics[width=\textwidth,trim=2.5cm 0 2.5cm 0, clip]{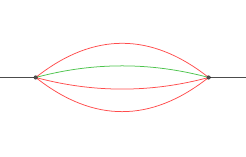}
         \caption{Class II}
         \label{fig:sampleII}
    \end{subfigure}
    \quad
    \begin{subfigure}[c]{0.3\textwidth}
         \centering
         \includegraphics[width=\textwidth,trim=2cm 0 2cm 0, clip]{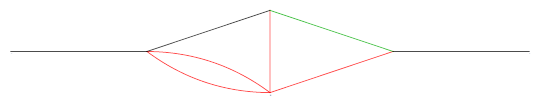}
         \caption{Class III}
         \label{fig:sampleIII}
    \end{subfigure}
    \caption{Sample of master integrals. Black, red and green lines denoted massive propagators with mass equal to $m_b$, $m_c$ and $ \sqrt{q^2}$. Dashed lines are massless propagators.
    The master integrals can have cuts through only one charm propagator~(a), 
    only three charm propagators~(b) and both kind of cuts~(c).}
    \label{fig:classes}
\end{figure}

For the decay rate, we need only the imaginary parts of the master integrals, i.e.\ the sum of 
all possible cuts.
Before discussing their solution, it is useful to divide them into three classes according to their cuts:
\begin{enumerate}[label=(\roman*)]
\item master integrals with cuts only through one charm quark propagator (and the fake particle with mass $q^2$). A sample integral is shown in Fig.~\ref{fig:sampleI}.
\item integrals where the cuts go through three charm propagators (see Fig.~\ref{fig:sampleII}).
\item integrals with both kind of cuts, i.e.\ one and three charm cuts (see Fig.~\ref{fig:sampleIII}).
\end{enumerate}
We would like to bring the system of differential equations in canonical form (or $\epsilon$-form)~\cite{Henn:2013pwa} and express 
the master integrals in terms of GPLs. Class II contains integrals beyond GPLs such as the sunrise diagram in Fig.~\ref{fig:sampleII} with two unequal masses. 

However, as observed already for the analytic calculation of the total rate at NNLO presented in Ref.~\cite{Egner:2023kxw}, the contribution given by cuts through three charm quarks can be neglected for realistic values of the ratio $m_c/m_b$.
It corresponds to the decay channel $b \to c \bar c c \ell \bar \nu_\ell$, which is rare and present only if $0 < \rho < 1/3$. 
For $m_c/m_b \simeq 0.25$, the branching ratio is very small, $O(10^{-7})$, because of the phase-space suppression and totally negligible compared to the current experimental accuracy.

Our strategy therefore is to solve the differential equations by considering only the cuts 
with one charm quark and neglecting the cuts with three charm quarks. To this end, we can effectively 
remove the masters in class II, while for class III we pick up only the one-charm cut contributions in the boundary conditions.

After such simplifications, the number of master integrals reduces to 87 and we proceed to find a basis transformation $\mathbb{T}(\rho, \hat q^2, \epsilon)$ such that the masters in the new basis $\bm J = \mathbb{T} \tilde{\bm J}$ satisfy a set of differential equations in canonical form.
We change variables from $\rho$ and $\hat q^2$ to
\begin{equation}
    x_\pm =
    \frac{1}{2}
    \left(
    1-\hat q^2+\rho^2 \pm \lambda^{1/2}(1,\hat q^2,\rho^2)
    \right),
    \label{eq::xpm}
\end{equation}
with $0 \le x_- \le x_+ \le 1$ and $\lambda(a,b,c) = a^2+b^2+c^2-2ab-2ac -2 bc$.
To find a rational transformation $\mathbb{T}$, we use \texttt{Libra}~\cite{Lee:2020zfb},
which implements the Lee’s algorithm~\cite{Lee:2014ioa}, in combination with \texttt{Fermatica}~\cite{fermatica}
to speed up the matrix transformations via the interface to the CAS program \texttt{Fermat}. 

As a first step, we find suitable transformations acting on the diagonal blocks that put them into $\epsilon$-form. 
At NLO, we observe that it is possible to bring the system of differential equations in canonical form when written in terms
of the variables $x_\pm$, however at NNLO this is not possible anymore.
The eigenvalues of the residue matrices of some blocks are of the form $a \epsilon \pm 1/2$, 
with $a$ an integer number.
Balanced transformations can only raise or lower the eigenvalues by an integer, so in order to bring such blocks 
into $\epsilon$-form, we need to apply an additional variable change. 
We find that all such eigenvalues with half-integers can be removed by switching from $x_+$ and $x_-$ to
the variables $u$ and $v$ defined by
\begin{align}
    \rho &= v, &
    \hat q^2 &= (1-uv) \left(1-\frac{v}{u}\right),
    \label{eqn:uv-definition}
\end{align}
with $0 < v \le u \le 1$.
However, we find it convenient not to perform the variable transformation globally 
because this would increase the degree of the poles in the residue matrices, 
making the reduction to $\epsilon$-form computationally more challenging.
Instead we take advantage of the \texttt{Notations} mechanism implemented in \texttt{Libra}.
The idea is to work with a system still expressed in terms of $x_+$ and $x_-$, 
but with the introduction of the notation
\begin{equation}
    u^2 = \frac{x_-}{x_+}.
\end{equation}
In each subblock, all variables  $x_+$, $x_-$ and $u$ appear. However
the dependence of the latter on $x_+$ and $x_-$ is always taken into account when 
computing derivatives and it is automatically simplified to first-order polynomial in $u$.
After bringing all diagonal blocks to $\epsilon$-form, \texttt{Libra} automatically reduces the off-diagonal blocks
to Fuchsian form and finds a suitable transformation independent ofn $x_+$ and $x_-$ to factorize out $\epsilon$. 
In the end, we bring the system of differential equation in canonical form:
\begin{equation}
    \mathrm{d} \bm \tilde {\bm J}(u,v,\epsilon) =
    \epsilon \, \mathrm{d} \mathbb{A}(u,v) \bm \tilde {\bm J}(u,v,\epsilon),
\end{equation}
where
\begin{equation}
    \mathbb{A}(u,v) = \sum_{i=1}^{14} A_i \log(\alpha_i).
\end{equation}
$A_i$ are $87 \times 87$ matrices with rational numbers
and $\alpha_i$ are the letters
\begin{align}
    \alpha_1 &= 1 - u, &
    \alpha_2 &= u, &
    \alpha_3 &= 1 + u, \notag \\
    \alpha_4 &= 1 - v, &
    \alpha_5 &=v, &
    \alpha_6 &= 1 + v, \notag \\
    \alpha_7 &= u - v, &
    \alpha_8 &= u + v, &
    \alpha_9 &= 1 - u v, \notag \\
    \alpha_{10} &=1 + u^2 - 2 u v,&
    \alpha_{11} &=1 + u^2 - u v, &
    \alpha_{12} &= 1 + u v, \notag \\
    \alpha_{13} &=u - v - u^2 v, &
    \alpha_{14} &=2 u - v - u^2 v.
\end{align}

Since all letters are linear in $v$, we write the solution of the differential equation in terms of 
GPLs in $v$ and letters that might depend on $u$, as well as Harmonic polylogarithms in $u$.

The boundary conditions to the differential equations are obtained using the auxiliary mass flow method \cite{Liu:2017jxz,Liu:2021wks} as implemented in \texttt{AMFlow} \cite{Liu:2022chg}.
We compute all 87 master integrals in four different kinematic points:
\begin{align}
 (q^2/m_b^2,m_c^2/m_b^2) \in \{(1/5,1/5),(4/25,4/25),(1/32,1/16),(1/16,1/32)\}~.
\end{align}
These points are sufficient to fix all boundary constants and at least have two additional kinematic points to check the resulting integrals.\footnote{Aside from integrals in class III, here the last two points allow for the three-charm cut and thus are not used.}
The master integrals are computed with sufficient numerical precision in order to obtain the boundary constants of the 
differential equations written in terms of transcendental numbers using the \texttt{PSLQ} algorithm~\cite{pslq}.

\section{Results}
\label{sec:results}
After the evaluation of the master integrals (one-charm cuts), we insert them into the
amplitude and perform the wave function and mass 
renormalization in the on-shell scheme~\cite{Melnikov:2000zc,Chetyrkin:2004mf,Marquard:2016dcn,Marquard:2018rwx,Fael:2020njb},
while we use $\overline{\mathrm{MS}}$ for the strong coupling constant.

Our main results are the analytic expressions for the functions $F_0, F_1$ and $F_2$ in the
differential decay rate in Eq.~\eqref{eqn:decayrate}. They are written in terms of
GPLs depending on $u$ and $v$, as defined in Eq.~\eqref{eqn:uv-definition}, which can be
evaluated numerically to high accuracy, e.g.\ with \texttt{GiNaC}~\cite{Bauer:2000cp} and \texttt{PolyLogTools}~\cite{Duhr:2019tlz}.
The explicit expressions for $F_0, F_1$ and $F_2$ are given as ancillary files~\cite{fael_2024_10781498}.

Moments of the $q^2$ spectrum are defined by
\begin{equation}
Q_n(q^2_\mathrm{cut}) = 
    \frac{1}{\Gamma_0}
    \int_{ q^2 > q^2_\mathrm{cut}}
     \,( q^2)^n \frac{\d \Gamma}{\d q^2} \, \d q^2,
\end{equation}
where $\Gamma_0 = G_F^2 m_b^5 |V_{cb}|^2/(192 \pi^3)$.
The moments can be expressed as series expansions in $\alpha_s$:
\begin{equation}
Q_n = \sum_{i \ge 0}
Q_n^{(i)} \left( \frac{\alpha_s(\mu_s)}{\pi}\right)^i.
\end{equation}
From the expressions for $F_i$, we calculate the coefficients in the perturbative
expansion via one-dimensional numerical integrations of the functions $F_i$:
\begin{align}
    \frac{Q_n^{(i)}(q^2_\mathrm{cut})}{m_b^{2n}} &= 
    \int_{ \hat q^2 > \hat q^2_\mathrm{cut}}
     \,( \hat q^2)^n F_i(\rho,\hat q^2) \, \d \hat q^2 
     \notag \\ &
     =
    \int_{u_\mathrm{min}}^1
    \left[ \left(1-u \rho \right) \left( 1-\frac{\rho}{u} \right) \right]^n
    \frac{(1-u^2)\rho}{u^2} F\Big(\rho , \hat q^2(\rho,u) \Big) \, \d u,
\end{align}
where 
\begin{equation}
    u_\mathrm{min} = 
    \frac{1}{2\rho}
    \left[
    1-\hat q^2_\mathrm{cut}+\rho^2
    -\lambda^{1/2}(1,\hat q^2_\mathrm{cut},\rho^2)
    \right],
\end{equation}
which reduces to $u_\mathrm{min} = \rho$ for $\hat{q}^2_\mathrm{cut} = 0$~.
We also define the normalized $q^2$ moments as
\begin{equation}
    \langle (q^2)^n \rangle_{q^2 \ge q^2_\mathrm{cut}} = 
    \frac{Q_n( q^2_\mathrm{cut})}{Q_0( q^2_\mathrm{cut})},
    \label{eqn:norm}
\end{equation}
and the centralized moments as
\begin{align}
    q_1( q^2_\mathrm{cut}) &= \langle q^2 \rangle_{q^2 \ge  q^2_\mathrm{cut}}, &
    q_{n}(q^2_\mathrm{cut}) &=
    \Big\langle \left(q^2-\langle q^2 \rangle \right)^n \Big\rangle_{q^2 \ge  q^2_\mathrm{cut}}
    \quad \mathrm{for}\, n\ge 2.
    \label{eqn:cent}
\end{align}
Moreover, with $\hat q_n$ we denote $\hat q_n = q_n/m_b^{2n}$.

\subsection*{On-shell scheme}
We now present our results for the centralized moments in the on-shell scheme.
We set the quark masses to $m_b^\mathrm{OS} = 4.6$~GeV and $m_c^\mathrm{OS} = 1.15$~GeV and 
numerically evaluate the coefficients in the perturbative expansions for $Q_n$, with $n=1,.\dots,4$.
Afterwards, we reexpand the ratios in Eqs.~\eqref{eqn:norm} and~\eqref{eqn:cent} in $\alpha_s$ up 
to second order. Our results for the moments without cuts ($q^2_\mathrm{cut}=0 \, \GeV^2$) read
\begin{align}
    \hat q_1 &=
    0.2185
    \left[
    1+0.1276 \left(\frac{\alpha}{\pi}\right) +0.4460  \left(\frac{\alpha}{\pi}\right)^2
    \right], \notag \\
    \hat q_2 &=
    0.02040
    \left[
    1+0.1382\left(\frac{\alpha}{\pi}\right)+0.9197  \left(\frac{\alpha}{\pi}\right)^2
    \right], \notag \\
    \hat q_3 &=
    1.1042 \times 10^{-3} 
    \left[
    1-0.2271\left(\frac{\alpha}{\pi}\right)+1.097 \left(\frac{\alpha}{\pi}\right)^2
    \right], \notag \\
    \hat q_4 &=
    8.895  \times 10^{-4} 
    \left[
    1 + 0.1677 \left(\frac{\alpha}{\pi}\right) + 1.591 \left(\frac{\alpha}{\pi}\right)^2
    \right],
\end{align}
which are in good agreement with the NNLO results presented in Ref.~\cite{Fael:2022frj}.
For a cut of $q^2_\mathrm{cut} = 3$ GeV$^2$, we obtain
\begin{align}
    \hat q_1 &=
    0.3022 \left[1 + 0.06894 \left(\frac{\alpha}{\pi}\right) + 0.3428 \left(\frac{\alpha}{\pi}\right)^2 \right], \notag \\
    \hat q_2 &=
    0.01151 \left[
    1 + 0.1433 \left(\frac{\alpha}{\pi}\right) +1.209 \left(\frac{\alpha}{\pi}\right)^2
    \right], \notag \\
    \hat q_3 &=
    5.1013 \times 10^{-4} \left[
    1 -0.2171 \left(\frac{\alpha}{\pi}\right) + 0.5447 \left(\frac{\alpha}{\pi}\right)^2
    \right], \notag \\
    \hat q_4 &=
    2.857 \times 10^{-4} \left[
    1 +0.1634 \left(\frac{\alpha}{\pi}\right) + 1.849 \left(\frac{\alpha}{\pi}\right)^2
    \right].
\end{align}

\subsection*{Kinetic scheme}
We discuss now the impact of higher-order QCD corrections to $q^2$ moments 
once a short-distance mass scheme is adopted for the quark masses.
We concentrate on the kinetic scheme employed in the global fits of Refs.~\cite{Gambino:2013rza,Alberti:2014yda,Bordone:2021oof,Bernlochner:2022ucr,Finauri:2023kte}.
In this scheme the on-shell mass of the bottom quark is replaced by 
the kinetic mass~\cite{Bigi:1996si,Czarnecki:1997sz,Fael:2020iea,Fael:2020njb} 
via the relation
\begin{equation}
m_b^\kin(\mu)= m_b^\OS-
  [\overline \Lambda(\mu)]_\pert
  -\frac{[\mu_\pi^2(\mu)]_\pert}{2m_b^\kin(\mu)} -
  O\left( \frac{1}{(m_b^\kin)^2} \right),
\end{equation}
while the charm quark mass is converted to the  $\overline{\mathrm{MS}}$ scheme.
At the same time, we include the contribution from power corrections to the 
moments up to $1/m_b^3$ (the relevant expressions can be retrieved from~\cite{Fael:2018vsp,Mannel:2021zzr}).
In the kinetic scheme, we redefine the HQE parameters $\mu_\pi^2$ and $\rho_D^3$ 
in the following way:
\begin{align}
  \mu_\pi^2(0) &= \mu_\pi^2(\mu)-[\mu_\pi^2(\mu)]_\pert, &
  \rho_D^3(0) &= \rho_D^3(\mu)-[\rho_D^3(\mu)]_\pert.
  \label{eqn:HQEParametersRedefinition}
\end{align}
Note that $\mu_\pi^2$ drops out for centralized $q^2$ moments, 
leaving a dependence only on $\rho_D^3$.
The perturbative version of $\mu_\pi^2$ and $\rho_D^3$ up to $\mathcal{O}(\alpha_s^3)$ can be found in the Appendix of Ref.~\cite{Fael:2020njb}.
The Wilsonian cutoff $\mu$ plays the role of scale separation between the short- and long-distance regimes in QCD.

In order to present our benchmark predictions of the $q^2$ moments for validation and comparison, 
we report the series expansion for centralized moments with $q^2_\mathrm{cut} = 0 \,\, \GeV^2$ and
$q^2_\mathrm{cut} = 4 \,\, \GeV^2$.
We adopt the HQE parameter definitions employed in Refs.~\cite{Benson:2003kp,Alberti:2014yda,Bordone:2021oof} (the so-called \textit{perp} basis).
We use scheme (A) as defined in Ref.~\cite{Fael:2022frj}:
in a first step the expressions for centralized moments are obtained in the on-shell
scheme where we retain terms up to $O(\alpha_s^2)$ at partonic level ($1/m_b^0$) while 
we discard higher QCD corrections for the subleading power corrections. 
Afterwards we apply the transition to the kinetic scheme.

We set the renormalization scale of the strong coupling constant
$\mu_s=m_b^\kin$ and use $\alpha_s^{(4)}(m_b^\kin)$ as expansion parameter,
i.e.\ we decouple the bottom quark from the running of $\alpha_s$, and we
reexpand the leading $1/m_b$ term in $\alpha_s^{(4)}$ up to second order.  
We use the input values
\begin{align}
  m_b^\kin(1\, \GeV) &=4.526 \, \GeV, &
  \overline m_c (3\, \GeV) &=0.989 \, \GeV, \notag \\
  \mu & = 1 \, \GeV , &
  \alpha_s^{(4)}(m_b^\kin) &= 0.2186.
\end{align}
In the following we present the results for the various contribution at leading order 
in the $1/m_b$ expansion for two different values of $q^2_\mathrm{cut}$.
We do not report the contribution from power suppressed terms,
however the terms originating from $[\rho_D^2(\mu)]_\mathrm{pert}$ are included.
Our results for the moments for $q^2_\mathrm{cut}=0$ GeV$^2$ read
\begin{align}
\hat q_1 & = 
0.2329 \left[1-0.1524 \api-1.791 \api^2\right], \notag \\
\hat q_2^2 &=
0.02353 \left[1-0.516 \api-4.474 \api^2\right], \notag \\
\hat q_3 &=
0.001451 \left[1-1.007 \api-7.408 \api^2\right], \notag \\
\hat q_4 &=
0.001202 \left[1-0.8404 \api-8.724 \api^2\right].
\label{eqn:q2kin0GeV2}
\end{align}
In case we apply a cut of $q^2_\mathrm{cut} = 4 \, \GeV^2$ we obtain
\begin{align}
\hat q_1 & = 
0.3503 \left[1-0.1628 \api-1.437 \api^2 \right], \notag \\
\hat q_2 & = 
0.01102 \left[ 1-0.7408 \api-6.236 \api^2 \right], \notag \\
\hat q_3 & = 
0.0005113 \left[ 1+0.06583 \api+4.269 \api^2\right], \notag \\
\hat q_4 & = 
0.0002659 \left[ 1-0.5384 \api-11.69 \api^2 \right].
\label{eqn:q2kin3GeV2}
\end{align}
We performed a comparison also with the corrections of order $\alpha_s^2 \beta_0$ 
(the so-called BLM corrections~\cite{Brodsky:1982gc}) recently presented in Table~1 of Ref.~\cite{Finauri:2023kte} 
and found good agreement. From the knowledge of the complete NNLO corrections, 
we observe that in the kinetic scheme the non-BLM contribution to the moments at 
$\mathcal{O}(\alpha_s^2)$ has in general the opposite sign of the BLM contribution, 
and is of comparable size. We conclude that the BLM approximation tends
to overestimate the NNLO corrections, especially in case one uses $\overline{m}_c(3 \, \GeV)$
as reference mass for the charm quark.

\begin{figure}[htb]
    \centering
    \includegraphics[width=0.49\textwidth]{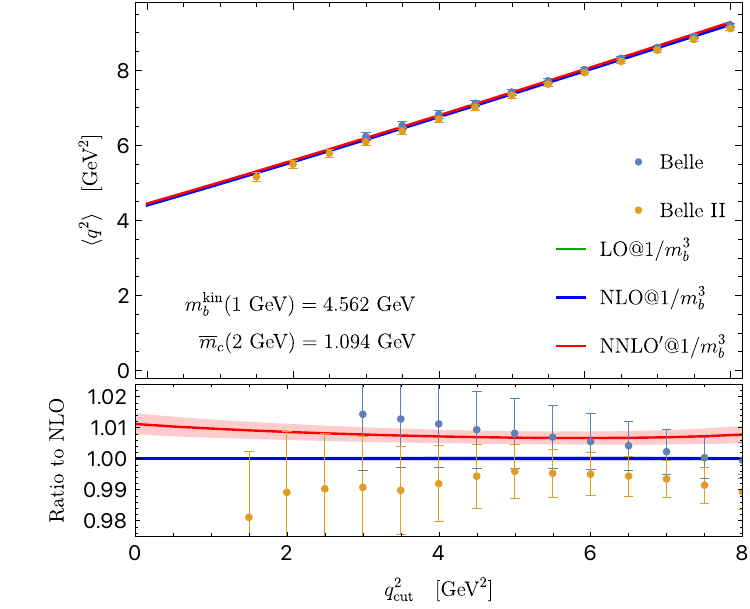}
    \includegraphics[width=0.49\textwidth]{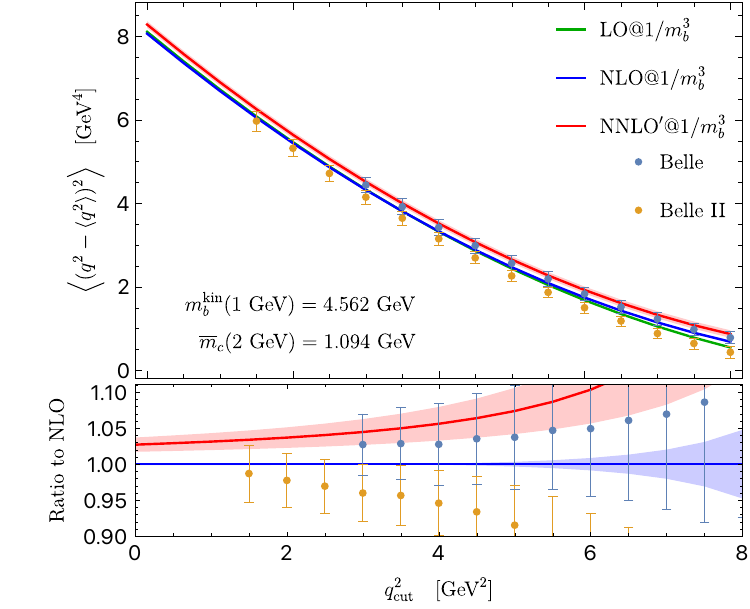}\\
    \includegraphics[width=0.49\textwidth]{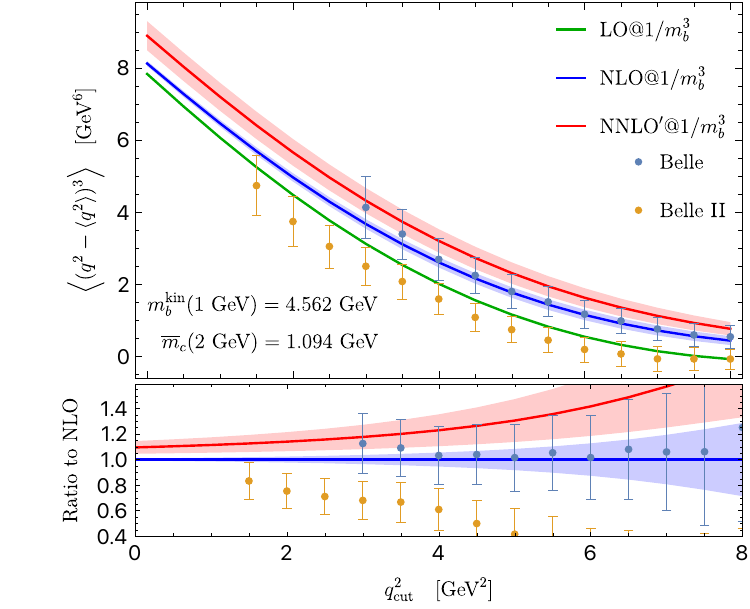}
    \includegraphics[width=0.49\textwidth]{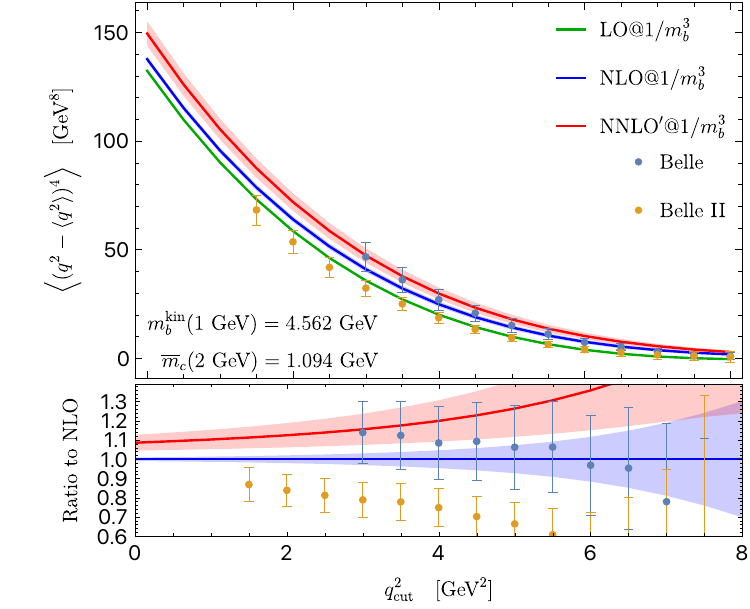}
    \caption{The first four $q^2$ moments of $B \to X_c \ell \bar \nu_\ell$ as a function of the lower cut $q^2_\mathrm{cut}$.
    The heavy quark masses are $m_b^\mathrm{kin}(1 \GeV) = 4.562 \, \GeV$ and $\overline{m}_c(2\, \GeV) = 1.094 \, \GeV$.
    For the HQE parameter we adopt the RPI basis up to $1/m_b^3$~\cite{Mannel:2018mqv,Fael:2018vsp} and 
    values from the fit in Ref.~\cite{Bernlochner:2022ucr}.
    Measurements from Belle~\cite{Belle:2021idw} and Belle II~\cite{Belle-II:2022evt}.
    }
    \label{fig:moments_mc2GeV}
\end{figure}

\begin{figure}[htb]
    \centering
    \includegraphics[width=0.49\textwidth]{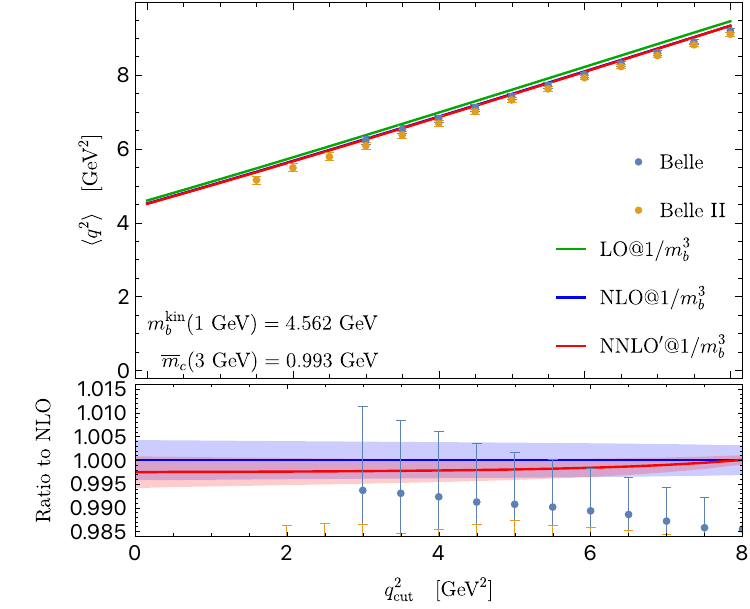}
    \includegraphics[width=0.49\textwidth]{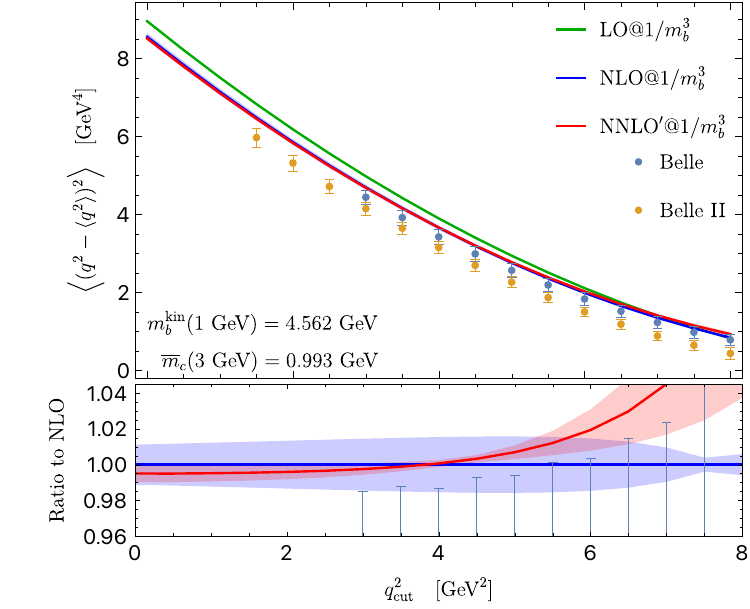}\\
    \includegraphics[width=0.49\textwidth]{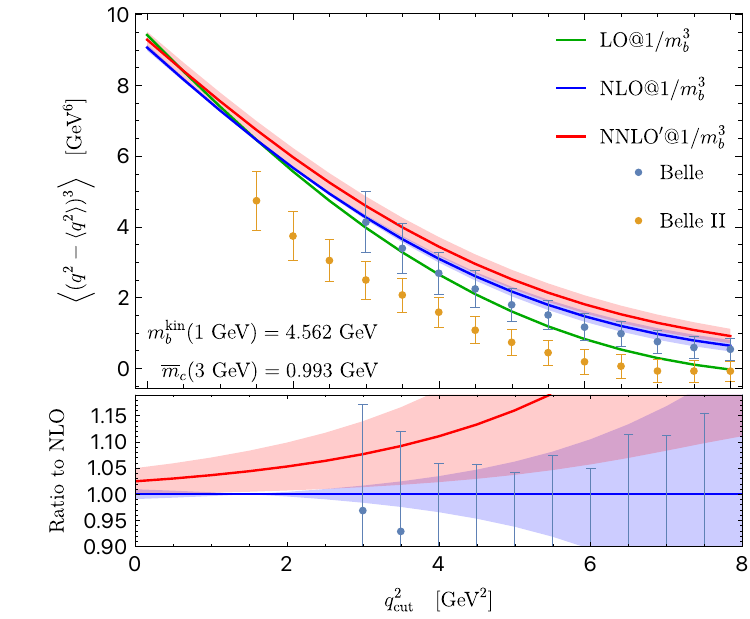}
    \includegraphics[width=0.49\textwidth]{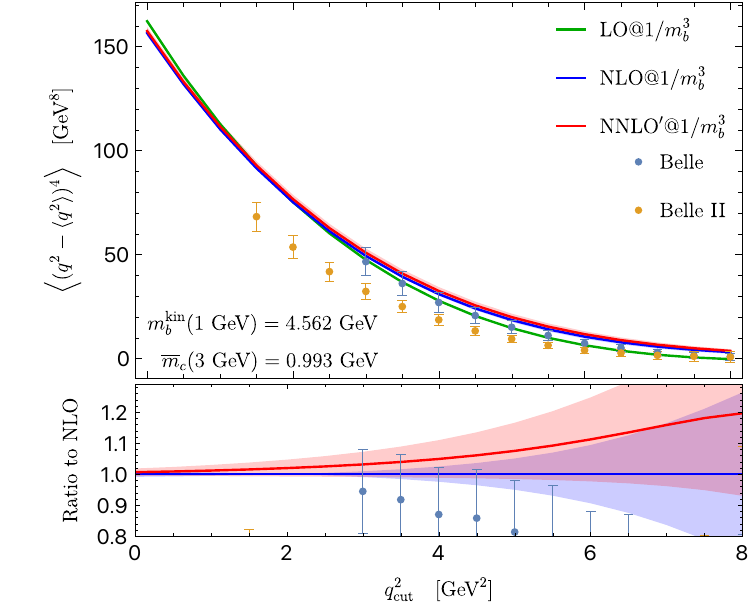}
    \caption{The first four $q^2$ moments of $B \to X_c \ell \bar \nu_\ell$ as a function of the lower cut $q^2_\mathrm{cut}$.
    The heavy quark masses are $m_b^\mathrm{kin}(1 \, \GeV) = 4.562 \, \GeV$ and $\overline{m}_c(3 \, \GeV) = 0.993 \, \GeV$.
    For the HQE parameter we adopt the RPI basis up to $1/m_b^3$~\cite{Mannel:2018mqv,Fael:2018vsp} and 
    values from the fit in Ref.~\cite{Bernlochner:2022ucr}.
    Measurements from Belle~\cite{Belle:2021idw} and Belle II~\cite{Belle-II:2022evt}.
    }
    \label{fig:moments_mc3GeV}
\end{figure}

Let us now discuss the size of the NNLO corrections and the impact on the 
global fits for $V_{cb}$.
In Figures~\ref{fig:moments_mc2GeV} and \ref{fig:moments_mc3GeV} 
we show our results for the first four centralized moments as a function of $q^2_\mathrm{cut}$. 
The predictions are compared with the Belle and Belle II measurements~\cite{Belle:2021idw,Belle-II:2022evt}.
At variance with the numerical values given in Eqs.~\ref{eqn:q2kin0GeV2} and~\ref{eqn:q2kin3GeV2},
in the plots we adopt the \textit{RPI} basis from Refs.~\cite{Mannel:2018mqv,Fael:2018vsp} 
and the values from the fit in Ref.~\cite{Bernlochner:2022ucr}
for the HQE parameters, $m_b$ and $m_c$. 

The green curves correspond to the LO prediction with power corrections
up to $1/m_b^3$. The blue curves include QCD NLO correction up to $1/m_b^3$, where we also include 
the $\alpha_s$ corrections to $\mu_G^2$ and $\rho_D^3$ calculated in Ref.~\cite{Mannel:2021zzr}.
The red curves, compared to the blue ones, additionally include the NNLO corrections at leading order in $1/m_b$ 
calculated in this article, which are denoted by NNLO' in the plots. 
The error bands are obtained by varying the renormalization scale in the range $m_b^\mathrm{kin}/2 < \mu_s < 2m_b^\mathrm{kin}$
and choosing $\mu_s = m_b^\mathrm{kin}$ as reference scale for the central value. 
We do not show the parametric uncertainty stemming from the HQE parameters.
The lower panel in each plot shows the ratio between the prediction at NNLO and NLO.

In Figure~\ref{fig:moments_mc2GeV}, we show the moments obtained with charm mass at a scale of 2 GeV, 
which is the default choice in the fit in Ref.~\cite{Bernlochner:2022ucr}. 
We observe that the NNLO corrections shift the prediction for $q_1$ and $q_2$ by 
a few percent in the low $q^2_\mathrm{cut}$ range.
For the third and fourth moment the impact is larger and close to a 10-15\% effects. The relative contribution
at higher values of $q^2_\mathrm{cut}$ becomes larger since the LO central value tends to vanish close to the end point.
Note that the use of $\overline{m}_c(2 \, \GeV)$ leads to accidentally small corrections at $O(\alpha_s)$ for all 
the moments. For $q_1$ and $q_2$ one can observe the overlap between the blue and green bands, while the red lines are much more separated.
Consequently, scale-variation alone does not provide a reliable uncertainty estimate for the NLO prediction.
In fact in this approximation, the scale uncertainty comes only from the variation of $\alpha_s$.
Since $\alpha_s$ is multiplied by a small number in case one uses $\overline{m}_c(2 \, \GeV)$, 
a rather small uncertainty is obtained.
Improving the prediction from the NLO to the NNLO, we observe that the $O(\alpha_s^2)$ coefficient
is not suppressed anymore and therefore NNLO error bands become larger than the NLO ones.

A better behaviour of the perturbative series is observed utilizing instead $\overline{m}_c(3 \, \GeV) = 0.993 \, \GeV$. 
The results are presented in Figure~\ref{fig:moments_mc3GeV}. Notice that now the $O(\alpha_s^2)$ corrections
are smaller than the $O(\alpha_s)$ ones, indicating a much better behaved expansion.
For the first and second moment the scale uncertainty is reduced from NLO to NNLO and the error bands
overlap. 
Even though the error bands overlap for the third and fourth moment, we still observe a larger uncertainty at NNLO.
This can be explained by the fact that the third and fourth moments, especially at high values of $q^2_\mathrm{cut}$,
receive sizable contributions form the power suppressed terms, which include perturbative corrections only up to NLO.
The uncertainty is not reduced in this case because cancellation of the $\mu_s$ dependence at partonic level is spoiled
by the lower accuracy in the perturbative expansion of the $1/m_b^2$ and $1/m_b^3$ terms.

Another notable effect observed in Fig.~\ref{fig:moments_mc2GeV} is that after inclusion of the NNLO corrections
the curves move to values higher than the experimental data points.
This does not indicate a tension between data and theory. In fact, since we use 
the HQE parameters from a fit accurate only up to NLO at partonic level~\cite{Bernlochner:2022ucr},
we naturally expect the NLO curves to show better agreement with the data. Notice that the blue curves include
also the NLO corrections at $1/m_b^2$ and $1/m_b^3$, which was not the case in Ref.~\cite{Bernlochner:2022ucr}.
The major effect in global fits after including the NNLO corrections would be a change of the favoured values
of the HQE parameters in order to shift downwards the red curves and accommodate the predictions with the data. 
In particular, since $\rho_D$ has a major impact in the $q^2$ moments and enters with negative coefficients, 
a fit with NNLO corrections would prefer higher values for $\rho_D$ compared to Ref.~\cite{Bernlochner:2022ucr}.

\begin{figure}[htb]
    \centering
    \includegraphics[width=0.49\textwidth]{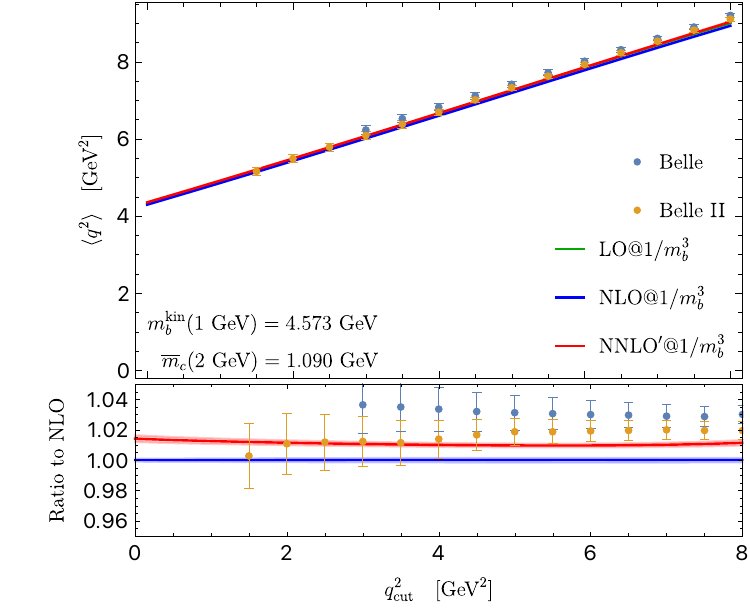}
    \includegraphics[width=0.49\textwidth]{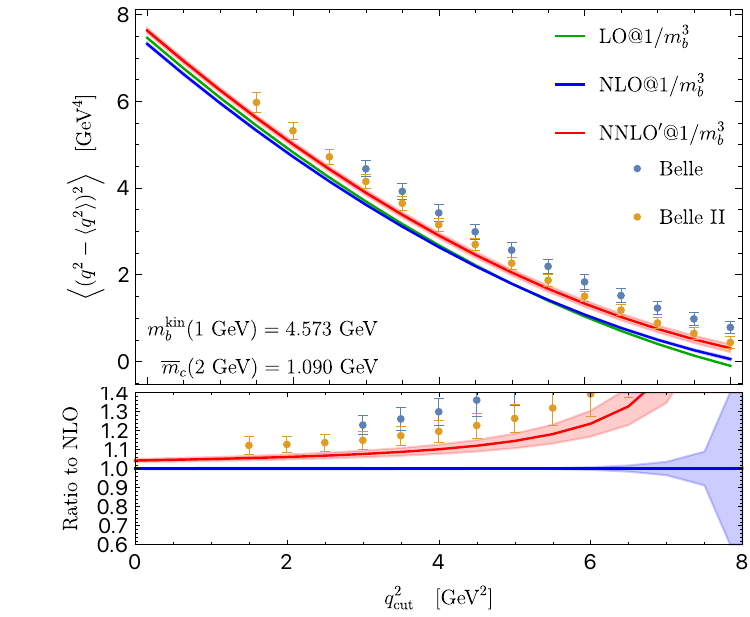}\\
    \includegraphics[width=0.49\textwidth]{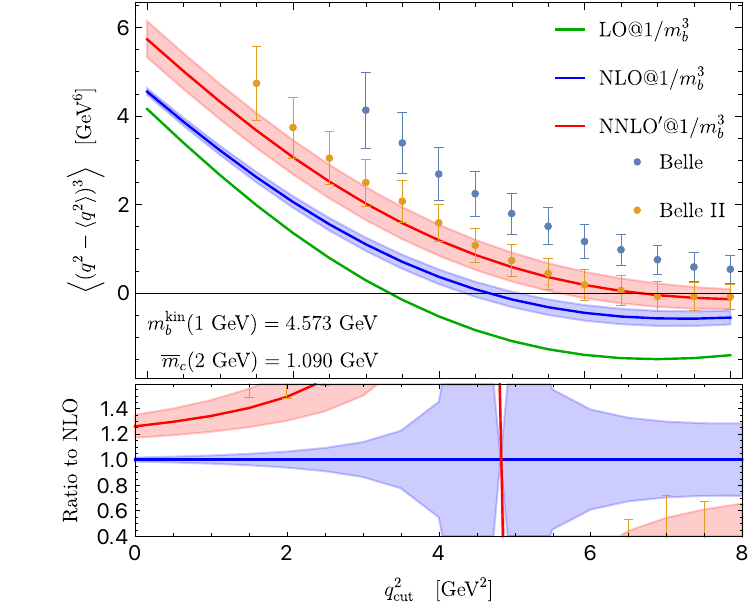}
    \includegraphics[width=0.49\textwidth]{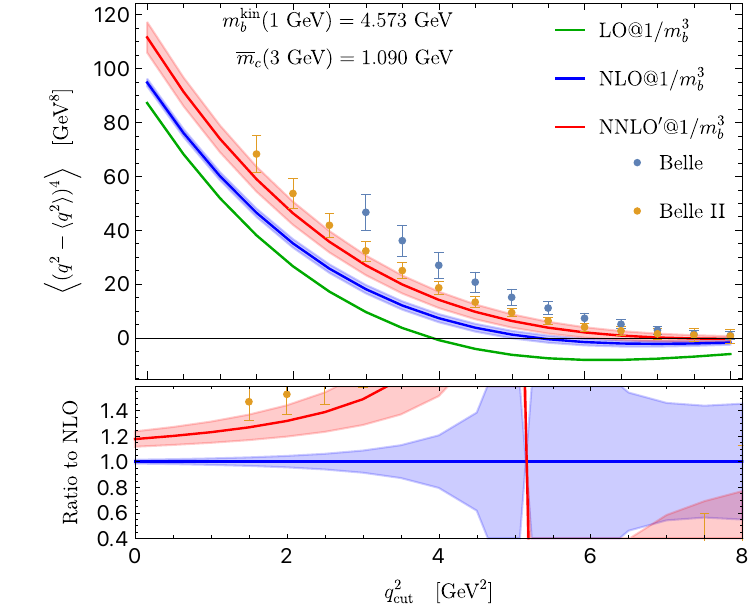}
    \caption{The first four $q^2$ moments of $B \to X_c \ell \bar \nu_\ell$ as a function of the lower cut $q^2_\mathrm{cut}$.
    The heavy quark masses are $m_b^\mathrm{kin}(1 \, \GeV) = 4.573 \, \GeV$ and $\overline{m}_c(2 \, \GeV) = 1.092 \, \GeV$.
    For the HQE parameter we adopt the \text{perp} basis up to $1/m_b^3$~\cite{Benson:2003kp} and
    values from the fit in Ref.~\cite{Finauri:2023kte}. The central values are obtained for a renormalization scale $\mu_s=m_b^\mathrm{kin}/2$.
    Measurements from Belle~\cite{Belle:2021idw} and Belle II~\cite{Belle-II:2022evt}.
    }
    \label{fig:moments_mc2GeV_perp}
\end{figure}

In addition to the analysis of the moments in the \textit{RPI} basis, we analyzed also the prediction 
with the fit setup from Ref.~\cite{Finauri:2023kte} and using the \textit{perp} basis
for defining the HQE parameters. We reached similar conclusions for what concerns the use of a charm mass at 
a scale of 2~GeV or 3~GeV: a charm mass at 3~GeV yields a better behaviour of the pertubative series
while a 2~GeV charm mass underestimates uncertainties at NLO.
For completeness, we report in Fig.~\ref{fig:moments_mc2GeV_perp} our results for a charm mass $\overline{m}_c(2 \, \GeV) = 1.092\,\GeV$, the default scheme in Ref.~\cite{Finauri:2023kte}.
We also mention that since the $\alpha_s^2 \beta_0$ corrections overestimates the 
$\alpha_s^2$ corrections, the NNLO predictions shown by the red curves in Fig.~\ref{fig:moments_mc2GeV_perp} 
lie below the experimental data. A curve showing the $q^2$ moment prediction with only the $\alpha_s^2 \beta_0$ corrections
would appear above the red curves, more in agreement with data.
We conclude that the inclusion of the complete NNLO corrections in the fit of~\cite{Finauri:2023kte} 
would bring the red curves upwards, towards the experimental data, preferring a lower value for $\rho_D$ (also 
in the \textit{perp} basis the coefficient of $\rho_D$ is negative).

\subsection*{Decay into a massive tau lepton}
While the $\mathcal{O}(\alpha_s^2)$ contributions to the $q^2$ spectrum are most relevant for the decay into light leptons, our expressions also apply to inclusive $b\rightarrow X_c \tau \bar{\nu}_\tau$ decays. The first measurement of the ratio
\begin{align}
    R(X) = \frac{\Gamma_{B\rightarrow X \tau \bar{\nu}_\tau}}{\Gamma_{B\rightarrow X \ell \bar{\nu}_\ell}}
\end{align}
was recently performed by the Belle II experiment \cite{Belle-II:2023aih}.
The current level of experimental precision is severely limited by systematic uncertainties related to the modelling of $B\rightarrow X \tau/\ell \bar{\nu}_{\tau/\ell}$ decays. However, recent progress in the description of $B$ meson decays into excited charm meson states \cite{Gustafson:2023lrz} will allow to address this issue in a data-driven way in the future, making the inclusion of $\mathcal{O}(\alpha_s^2)$ corrections relevant.

In the on-shell scheme, our results for the integrated $b\rightarrow X_c \tau \bar{\nu}_\tau$ decay rate without cut on $q^2$ agrees with Ref.~\cite{Biswas:2009rb}. In the kinetic scheme, with 
$m_b^\mathrm{kin}(1 \, \GeV) = 4.526\,\GeV$, $\overline{m}_c(3 \, \GeV) = 0.993 \, \GeV$ and at leading order in $1/m_b$, we obtain
\begin{align}
    R(X_c) = 0.241\left[1 - 0.156 \api - 1.766 \api^2\right]~.
\end{align}
The perturbative convergence is similar to the case of the $q^2$ moments.
This prediction can be improved by incorporating  $1/m_b$ suppressed terms at LO and NLO \cite{Moreno:2022goo}.

Our results allow to obtain predictions for $R(X_c)$ with a lower cut on $q^2$. For $q^2_\text{cut} = 6\,\text{GeV}^2$, we obtain
\begin{align}
    R(X_c)\Big|_{q^2 > 6\,\text{GeV}^2} = 0.350\left[1 - 0.782 \api - 8.355 \api^2\right]~.
\end{align}
The higher order corrections clearly become more relevant if a cut is introduced.

Furthermore, the ratio increases with increasing $q^2_\text{cut}$, as terms proportional to $m_\tau^2/q^2$ and phase-space effects become less relevant. Consequently, it could be advantageous to perform measurements of $R(X)$ using a lower cut on $q^2$ in the future, as it enriches the fraction of $B\rightarrow X\tau\bar{\nu}_\tau$ decays, rejects backgrounds and cuts away the regions where the $B\rightarrow X\ell\nu$ modelling is most problematic. In addition a lower cut on $q^2$
allows for the improved inclusion of momentum requirements on signal leptons due to detector thresholds and the reduction of uncertainties associated to the modelling of final state radiation \cite{Herren:2022spb}.
Most of the analysis strategy of the Belle measurement of the $q^2$ moments \cite{Belle:2021idw} could thus carry over to a future measurement of $R(X)$, with the exception of a cut on the difference of the missing energy and the missing momentum in a given event.\footnote{The main strategy to improve the $q^2$ resolution in the Belle II measurement \cite{Belle-II:2022evt} can not be applied to semitauonic decays as it depends on the presence of exactly one neutrino in the event.}

\section{Conclusions}
\label{sec:conclusions}
In this article we presented the complete NNLO QCD corrections to the $q^2$ spectrum of inclusive semileptonic $B$ decays.
The differential rate with respect to the leptonic invariant mass $q^2$ is obtained by calculating the 
imaginary part of the $b \to b$ 2-point function in the presence of a constraint on $q^2$, which can be implemented 
in a convenient way by replacing the charged lepton-neutrino loop with a fake particle with mass $q^2$.
After reduction to master integrals, we leverage the method of differential equations to calculate the decay rate analytically. 
To this end, we restricted the calculation to the cuts through only one charm line which allowed us to 
bring the system of differential equations in canonical form and express the master integrals in terms of GPLs.

In light of the recent measurements of Belle and Belle II of the $q^2$ moments, we studied the impact of the 
NNLO corrections on the moments as a function of $q^2_\mathrm{cut}$. 
We observe that the $\mathcal{O}(\alpha_s^2)$ corrections are sizable especially for the default choice of the 
charm mass $\overline{m}_c(2 \, \GeV)$ in the global fits of~\cite{Bernlochner:2022ucr,Finauri:2023kte}. 
The relative ratio to the NLO prediction reaches the 1-5\% level, depending on the cut on $q^2$, while  
for the higher moments it is larger and of about 10-20\%. 
For a charm mass at the scale of $\overline{m}_c(3 \, \GeV)$ we observe a better behaviour 
of the perturbative series, with the expected reduction of theoretical uncertainties when 
including the $\mathcal{O}(\alpha_s^2)$ corrections.
We applied our results also to the decay into a tau lepton, and proposed a measurement of $R(X_c)$ 
using a lower cut on $q^2$ to enrich the fraction of $ B \to X_c \tau \bar\nu_\tau$ events.
We provide our results for the $q^2$ spectrum in electronic form as ancillary files.
They can be employed to incorporate the NNLO corrections into global fits of inclusive semileptonic
B-decays, in particular to take advantage of the recent measurements of $q^2$ moments by Belle and Belle II.

\section*{Acknowledgements}  

We thank J.\ Usovitsch and F.\ Lange for the help with {\tt Kira} and 
R.\ N.\ Lee for the help with {\tt Libra}.
We thank also P.\ Gambino, R.\ van Tonder and K.\ Vos for discussion and correspondence.
The work of M.F.\ is supported by the European Union’s Horizon 2020 
research and innovation program under the Marie Sk\l{}odowska-Curie grant agreement 
No.\ 101065445 -- PHOBIDE.
This research was supported in part by the Swiss National Science Foundation (SNF) under contract 200021-212729.

\bibliographystyle{jhep} 
\bibliography{main.bib}

\end{document}